# PubMed Labs: An experimental platform for improving biomedical literature search

Nicolas Fiorini[1], Kathi Canese[1], Rostyslav Bryzgunov[1], Ievgeniia Radetska[1], Asta Gindulyte[1], Martin Latterner[1], Vadim Miller[1], Maxim Osipov[1], Michael Kholodov[1], Grisha Starchenko[1], Evgeny Kireev[1], Zhiyong Lu[1,*]

[1]National Center for Biotechnology Information (NCBI), National Library of Medicine (NLM), National Institutes of Health (NIH), 8600 Rockville Pike, Bethesda, MD 20894 USA

*Correspondence: zhiyong.lu@nih.gov

## Abstract

PubMed is a freely accessible system for searching the biomedical literature, with approximately 2.5 million users worldwide on an average workday. We have recently developed PubMed Labs (www.pubmed.gov/labs), an experimental platform for users to test new features/tools and provide feedback, which enables us to make more informed decisions about potential changes to improve the search quality and overall usability of PubMed. In doing so, we hope to better meet our user needs in an era of information overload. Another novel aspect of PubMed Labs lies in its mobile-first and responsive layout, which offers better support for accessing PubMed on the increasingly popular use of mobile and small-screen devices. Currently, PubMed Labs only includes a core subset of PubMed functionalities, e.g. search, facets. We encourage users to test PubMed Labs and share their experience with us, based on which we expect to continuously improve PubMed Labs with more advanced features and better user experience.

## Introduction

As the biomedical literature grows at an exponential rate, the National Center for Biotechnology Information (NCBI) at the National Library of Medicine has recently developed PubMed Labs (www.pubmed.gov/labs), an experimental platform for users to test new features/tools and provide feedback, which enables us to make more informed decisions about potential changes to improve the search quality and overall usability of PubMed (www.pubmed.gov). PubMed Labs is implemented as a standalone service, separate from the production operation of PubMed (1), so that it is non-disruptive to the routine information seeking process of our current PubMed users. Furthermore, as an experimental platform, it provides opportunities to experiment and innovate within a powerful sandbox to advance the limits of literature search and gather user feedback.

PubMed Labs has several unique features that distinguish it from PubMed and other search systems for the biomedical literature (2-4): (i) By default, given a free-text query as input, search results are sorted by Best Match[1] in order to provide users with the most pertinent information (in PubMed, the default sort order is Most Recent) as suggested by previous research (5-8). This is based on a newly developed cutting-edge relevance search algorithm using machine learning. In addition, to help users identify articles of interest, search results include snippets, useful highlighted text fragments from the article abstract that are selected based on their relatedness with the user query. (ii) PubMed Labs has a more modern user interface. Users will find it easier to discover related content (e.g. similar articles, references, and citations). (iii) Compatibility with smaller screen portable devices (e.g. phones, tablets and laptops) is optimized to ensure the best possible searching and reading experience on such devices; (iv) Finally, please note that by design PubMed Labs includes only a limited set of highly used features (9), and not the entire set currently available in PubMed. Based on public testing and feedback, we will iteratively include additional functions and improve the system towards a next-generation PubMed (10) over time. PubMed Labs was first made public in October 2017 and is currently accessed by thousands of users from around the world each day.

## Materials and Methods
### Data Indexing
As of 2018, there are over 28 million articles in PubMed Labs where each article is indexed via separate data fields: titles, abstracts, MeSH terms, etc. Different from PubMed, we use Solr, an open-source enterprise search platform (http://lucene.apache.org/solr/), for document indexing and retrieval. In addition to its scalability and reliability, Solr also provides many out-of-the-box search functionalities, such as better understanding wildcards ("*"). For example, because PubMed limits the number of variants for wildcards, the query therap* and cell[jour] and 2017[year] only yields 77 hits in PubMed while 129 results are returned by Solr in PubMed Labs[2]. Another notable Solr feature is to integrate synonyms during term indexing such that it results in significant improvements in search time. Finally, in PubMed Labs, the document data for indexing is newly generated by merging content from PubMed, Books, and PubMed Central (PMC) such that it allows the display of relevant information not available in PubMed (e.g. references from PMC).

---

[1] https://www.nlm.nih.gov/pubs/techbull/jf17/jf17_pm_best_match_sort.html
[2] Accessed on February 23, 2018

### User Interface (UI) Infrastructure
PubMed Labs is a Django (https://www.djangoproject.com/) application on the front-end, making use of the latest Web technologies and standards. It is compatible on any screen size and provides a fresh and consistent look-and-feel throughout the application.

### Integration of Third-Party Analytics Tools
PubMed Labs is first and foremost an experimental platform, therefore seeking and analysing user feedback is a critical component for the system. To this end, we used a third-party analytic tool in order to gather aggregated user behaviours and trends. This provides a convenient way for us to investigate the utility of certain features, and to determine which ones are more needed and vice versa. Additionally, we use such a tool to set up A/B tests, a controlled experiment for comparing variants of certain features.

## How to Use PubMed Labs
### How to Search
In order to run a search, users can type their queries made of free keywords in the search box (see Fig 1a). As in PubMed, field tags (e.g. "[author]") can be attached to the queries with the same syntax conventions[3]; Boolean operators (e.g. "AND", "OR", "NOT") are supported and the query syntax remains the same. For sort orders, PubMed Labs currently supports the two most used ones in PubMed (i) Best Match and (ii) Most Recent. By default, results are retrieved using the Best Match sort order as it aims to return the most relevant information given a query. The Best Match algorithm is built on a state-of-the-art machine learning approach and incorporates many relevance signals to find the most pertinent information at the top of the returned results. Meanwhile, some user needs may be better served with the Most Recent sort order (e.g. browsing the latest issue of a journal). Thus, in PubMed Labs the two sort orders are displayed next to each other and can be switched easily (see Fig 1b) with a single click (PubMed Labs also remembers the last sort order users choose to use). We believe the single-click switch is also convenient for users to compare results provided by both sort orders in some use cases. Note that query auto completion and related searches are also available (see Fig 1c and Fig 1d) as in PubMed.

---

[3] https://www.ncbi.nlm.nih.gov/books/NBK3827/#pubmedhelp.Search_Field_Descriptions_and

**Figure 1**. PubMed Labs search results page with highlighted features. (a) Search box. (b) Sort order toggle. (c) Query auto completion. (d) Search facets. (e) Highlighted search terms in title and snippet. (f) Related searches.

### How to Examine Search Results

If a search returns only one result, PubMed Labs displays the article abstract page, the same behaviour as PubMed. For all other searches, a summary of search results, is displayed (see Fig 1). This page is critical in all search engines, as it allows the users to quickly get an idea of the results returned. Hence, this page needs to provide enough information for users to judge which articles might satisfy their information need while not overwhelming them. For PubMed Labs, this page includes both traditional as well as new features.

*Article details and snippets.* In PubMed Labs, only the top 10 results are displayed in the results page, compared to 20 in PubMed. However, the matching query terms are highlighted for each result in PubMed Labs, to help users better understand why they were returned. This would be especially useful if an article matched a synonymous term in the query. For each returned article, this author list has been shortened for added consistency and readability in the results page. Up to two authors are listed, and articles with more than two authors are shown with the name of the first author, followed by "et al". However, if a query contains an author name (e.g. koonin ev), matching articles will always highlight the author name in the results.

The journal name and the article type (if it is a review or a clinical trial) are provided next to the author list. PubMed Labs also brings an important new feature by showing snippets, which are excerpts from abstracts that best match the query and provide additional contextual information. This helps to show how the returned article is related to the search query.

*Search facets.* As in PubMed, search facets are displayed on the left to enable users to easily refine their search (see Fig 1f) and narrow down search results. While this list is less comprehensive than its PubMed counterpart, these facets are the most used ones and should satisfy most of the needs while we work on adding more. Text availability and article type can be combined (e.g. by selecting both "Abstract" and "Full text", which will return articles matching any of the two), while only one publication date facet can be selected at a time. The selected filters are saved within a search sequence (i.e., during refinements and back and forth with articles). A reset button at the bottom allows users to conveniently remove all filters associated with the search with a single click.

## How to Examine Results of Each Individual Article

The abstract page, displayed after a click on an article title in the search results page, is another critical component of our system. It provides more details about the article and includes rich information for the user to make a decision in their next step (e.g. downloading the full text, refining the query, browsing similar articles, etc.).

**Figure 2**. PubMed Labs abstract page with highlighted content. (a) Major publication type. (b) Abbreviated journal name. (c) Publication date. (d) List of full author names. (e) Author affiliations. (f) PMID and PMCID. (g) Abstract. (h) Figures. (i) Full text links. (j) List of articles citing this paper in PMC. (k) References listed in this article. (l) Full list of publication types. (m) MeSH concepts indexing this article.

*Publication metadata.* The top of our abstract page is dedicated to the article's metadata. Particularly, as detailed in Fig 2, it provides (a) the publication type, (b) journal name, (c) publication date, (d) list of author names, (e) their affiliations, and (f) the article's PMID. The main difference is that we display full author names in PubMed Labs rather than just initials and last name as in PubMed.

*Abstract & figures.* Right underneath in the main area of the page lay the article contents, i.e. the abstract (Fig 2g) and figures (Fig 2h), when available. The figure display has been entirely revamped to provide a cleaner and more user-friendly look, and it features an in-page full-size view that you can share simply by copying the URL.

*Similar articles.* Another highly used feature in the PubMed abstract page is the Similar Articles. These suggestions are calculated for every article (11) and the most similar ones can provide important and interesting additional information. In PubMed Labs, the title, first author, journal name, publication year and first two lines of the abstract are displayed for the five most similar articles.

*Citation data.* Two types of references are displayed next: (i) cited by and (ii) article references. The former shows only articles available in PMC that are citing this article. The latter is the entire list of references with links to PubMed and PMC, when available. Both are conditioned by data availability, so not all articles have this information.

*Next/previous article.* As in PubMed Mobile, PubMed Labs now shows next and previous buttons at the very bottom of the page, displayed in the results page format. When used on small-screen devices (e.g. phones), PubMed Labs also shows these buttons at any scrolling position, with a quick glance at the articles when hovered over.

*Right column.* As usual, full text links are displayed at the top right position of the screen. Note, however, that this column is now made sticky, meaning that scrolling does not make it disappear. Moreover, a new Cite button is showed, which allows the users to have easy access to citation information in AMA, MLA and APA format. The citation can also be downloaded in RIS format, useful for citation managers. Finally, the right column features a navigation menu that allows users a quick glance at what associated content is available for the article.

## Use Cases and Usage Statistics

Since PubMed Labs includes nearly all features of PubMed Mobile and that it is responsive to small screen devices, we started inviting users on PubMed Mobile to try out PubMed Labs with a promotion banner displayed on the PubMed Mobile website. As of April 2018, on an average weekday there are over 3,000 users from around the globe with approximately 5,000 searches and 7,000 page views. Both sort orders are used by our users: Best Match (94%) and Most Recent (6%). As in PubMed, the most popular user activities after reading the abstract are to retrieve the full text and/or read similar articles.

Given its experimental nature, we have also started performing various a/b testing to refine the interface and certain features in PubMed Labs. As a simple example, we are currently comparing 4 variants for the newly created Cite button in the abstract page (see Fig 3). The basic version has the background colour in grey with the text "Cite". Alternatively, we are testing the combinations of another background colour (blue) as well as a different wording ("Cite article") for this. Preliminary results so far indicate that "Cite" with a blue background is the most preferred option by our users.

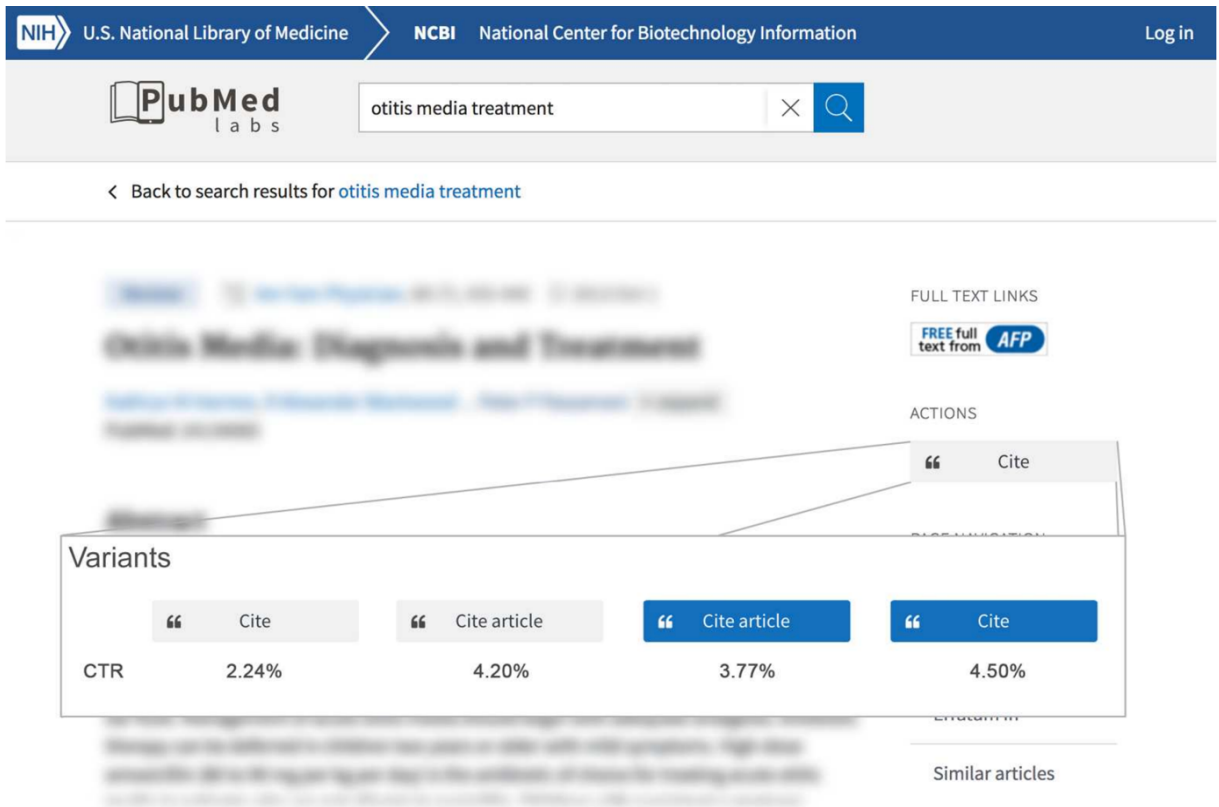

**Figure 3**. An experiment currently running on PubMed Labs, where variants of the cite button design are being tested, with their associated click-through rates.

## Summary and Future Directions

As mentioned, PubMed Labs currently includes a subset of core functionalities of PubMed (search, major sort orders, most used facets, etc.) as well as a number of new experimental features (e.g. snippets or the Cite button). Nonetheless, its functionality is limited compared to PubMed where it offers additional features such as advanced search, MyNCBI, etc. A future direction is to iteratively add and test new features based on user input. To that end, we have implemented a Feedback button on all PubMed Labs pages. We strongly encourage our users to test PubMed Labs and share their experience with us, which complements the insight we can glean from usage analysis and other user research. With the help from our users, we hope to jointly improve PubMed Labs with more features and better user experience in the future.

## Data Availability

This website is freely available to all via the web, no login or registration is required and not be password-protected.

## Acknowledgment

We would like to thank NCBI/NLM leadership for their support and many others (including our users) for their feedback and helpful discussion on the project. This research was supported by the Intramural Research Program of the NIH, National Library of Medicine.

## References

1. NCBI Resource Coordinators. Database resources of the National Center for Biotechnology Information. Nucleic acids research. 2015;43(Database issue):D6-17.
2. Levchenko M, Gou Y, Graef F, Hamelers A, Huang Z, Ide-Smith M, et al. Europe PMC in 2017. Nucleic acids research. 2018;46(D1):D1254-D60.
3. Doms A, Schroeder M. GoPubMed: exploring PubMed with the Gene Ontology. Nucleic acids research. 2005;33(Web Server issue):W783-6.
4. Lu Z. PubMed and beyond: a survey of web tools for searching biomedical literature. Database : the journal of biological databases and curation. 2011;2011:baq036.
5. Nourbakhsh E, Nugent R, Wang H, Cevik C, Nugent K. Medical literature searches: a comparison of PubMed and Google Scholar. Health information and libraries journal. 2012;29(3):214-22.
6. Siadaty MS, Shu J, Knaus WA. Relemed: sentence-level search engine with relevance score for the MEDLINE database of biomedical articles. BMC medical informatics and decision making. 2007;7:1.
7. Bui DD, Jonnalagadda S, Del Fiol G. Automatically finding relevant citations for clinical guideline development. Journal of biomedical informatics. 2015;57:436-45.
8. Eaton AD. HubMed: a web-based biomedical literature search interface. Nucleic acids research. 2006;34(Web Server issue):W745-7.
9. Islamaj Dogan R, Murray GC, Neveol A, Lu Z. Understanding PubMed user search behavior through log analysis. Database (Oxford). 2009;2009:bap018.
10. Fiorini N, Lipman DJ, Lu Z. Towards PubMed 2.0. eLife. 2017;6:epublish.
11. Lin J, Wilbur WJ. PubMed related articles: a probabilistic topic-based model for content similarity. BMC bioinformatics. 2007;8:423.